\documentclass[aps,showpacs,amsmath,tightenlines,preprintnumbers]{revtex4}
\usepackage{graphicx}

\newcommand{\mb}[1]{\mbox{\scriptsize #1}}
\newcommand{\mt}[1]{\mbox{\tiny #1}}
\newcommand{\sigmac}{\bar{\psi}\psi}
\newcommand{\pic}{\bar{\psi}i\gamma_5\vec{\tau}\psi}

\newcommand{\atan}{\mbox{arctan}}

\begin{document}
\preprint{RCNP-Th02019}

\title{Chiral phase properties of finite size quark droplets\\
in the Nambu--Jona-Lasinio model}

\author{O. Kiriyama}
\email{kiriyama@rcnp.osaka-u.ac.jp}
\author{A. Hosaka}
\email{hosaka@rcnp.osaka-u.ac.jp}
\affiliation{Research Center for Nuclear Physics, 
Osaka University, Ibaraki 567--0047, Japan}

\date{\today}

\begin{abstract}
Chiral phase properties of finite size hadronic systems are 
investigated within the Nambu--Jona-Lasinio model. 
Finite size effects are taken into account by making use of the 
multiple reflection expansion. 
We find that, for droplets with relatively small baryon numbers, 
chiral symmetry restoration is enhanced 
by the finite size effects. However the radius of the stable droplet 
does not change much, as compared to that without the 
multiple reflection expansion. 
\end{abstract}


\maketitle

\section{Introduction}
The behavior of (finite lumps of) quark matter is of 
great interests in cosmology, neutron stars, cosmic ray physics 
and heavy ion collisions \cite{EW,SQM}. 
Absolutely stable nonstrange quark matter contradicts ordinary nuclei 
consisting of neutrons and protons. However, the existence of stable 
strange quark matter is still an open question 
and it may be realized in the form of strangelets, 
small lumps of strange quark matter. 
The (meta)stability of nonstrange and strange quark matter has been 
investigated within the MIT bag model \cite{CEA}, 
quark mass density dependent model \cite{FRW}, 
Nambu--Jona-Lasinio (NJL) model \cite{NJL}, and so on \cite{OM}. 
In the MIT bag model, which assumes that 
asymptotically free quarks are confined in a bag, 
the bag constant and the current quark masses 
are phenomenological input parameters. 
Farhi and Jaffe \cite{FJ} found a reasonable range of these parameters 
in which strange quark matter is stable, while nonstrange quark matter 
is unstable as compared to a gas of $^{56}{\rm Fe}$. 
The quark mass density dependent model takes account of 
the confinement mechanism by introducing density dependent 
quark masses. Fowler {\it et al.} \cite{FRW} concluded that 
the critical density for a phase transition 
between nuclear and quark matter is sensitively 
dependent on the confinement mechanism. 
In the NJL model, the bag constant and constituent quark masses are generated 
dynamically in marked contrast to the above mentioned models. 
In Ref. \cite{MB}, it has been argued that stable nonstrange quark matter lies 
in chirally restored phase and, then, the model shows similar behavior 
to the MIT bag model (see, also, Ref. \cite{MA}). 
The stability of strange quark matter has been also studied 
and concluded that strange quark matter is not absolutely stable 
when realistic values of current quark masses and coupling constants 
are used \cite{BO}.

On the other hand, unlike bulk systems, finite size effects cannot be 
ignored in the study of finite lumps of quark matter. 
So far, the MIT bag model and Fermi gas model including finite 
size effects have been widely used 
to study their properties \cite{FJ,BJ,VGN,GEA,JM,GJ}. 
For finite quark matter, it is known that 
the lumps are not energetically favored by finite size effects. 
The disfavor also arises from large strange quark masses. 
Therefore, finite size effects on dynamical quark masses 
(i.e. the behavior of chiral symmetry) 
might have significant effects on their properties. 
In this paper, we adopt the NJL model 
that is a simple tractable model 
respecting chiral symmetry. 
Then we consider isospin symmetric 
($n_u=n_d$ with $n_i=\langle q_i^\dagger q_i\rangle$) quark droplets. 
Finite size effects are taken account of 
by making use of the multiple reflection expansion (MRE) 
\cite{MRE,JM,BJ} which is an approximation for the density of states 
in finite systems. 
The MRE itself has been derived for free massless particles or 
nonrelativistic particles. 
However, the MRE has been used to calculate 
the thermodynamic quantities of {\it massive} quarks 
and reproduced well the results of the MIT bag model 
\cite{SQM,JM,BJ,GJ}. 
For instance, it can be checked that the vector 
($\psi^{\dagger} \psi$) and scalar ($\bar \psi \psi$)
densities in the MIT bag model are very similar to those 
computed in the MRE with little $r$ dependence 
for bags of radius $R \agt$ a few fm.  
Therefore, we expect that the MRE also works in the NJL model 
in the mean field approximation with a constant $m$ inside 
the bag and apply the MRE to the effective potential of the model. 

This paper is organized as follows. 
In Sec. II, we formulate the effective potential of spherical quark 
droplets that consist of an equal number of up and down quarks. 
The Schwinger--Dyson equation and related thermodynamic quantities are 
also derived. 
In Sec. III, we present numerical results. 
Section IV is devoted to conclusions. 

\section{The model}
In this section, the effective potential and related 
thermodynamic quantities of spherical quark droplets 
are derived in the two--flavor NJL model. 
We consider up and down quarks to be massless. 
Then, the ${\rm SU(2)}_L \times {\rm SU(2)}_R$ chirally symmetric Lagrangian 
is given by
\begin{eqnarray}
{\cal L}=\bar{\psi}i\gamma^\mu\partial_\mu\psi
+G\left[(\sigmac)^2+(\pic)^2\right],
\end{eqnarray}
which is our starting point. 
Here, $\psi$ denotes a quark field with two flavors ($N_f=2$) and 
three colors ($N_c=3$), 
and $G$ is a dimensionful coupling constant ($\mbox{mass}^{-2}$). 
The Pauli matrices $\vec{\tau}$ act in the flavor space.

Let us first discuss the effective potential for bulk systems. 
In the mean-field (Hartree) approximation, the effective potential 
$\omega$ at finite temperature $T$ and 
quark chemical potential $\mu$ is written as \cite{NJL}
\begin{eqnarray}
\omega=\frac{m^2}{4G}
-\nu\int\frac{k^2dk}{2\pi^2}E_k-\nu T\int\frac{k^2dk}{2\pi^2}
\ln\left[1+e^{-\beta(E_k+\mu)}\right]
\left[1+e^{-\beta(E_k-\mu)}\right],
\end{eqnarray}
where $\nu=2N_fN_c$, $E_k=\sqrt{k^2+m^2}$, $\beta=1/T$, 
with $m$ being the dynamically generated (constituent) quark mass. 
Since the model is not renormalizable, we have to specify 
a regularization scheme. 
Throughout this paper, we use a sharp cutoff $\Lambda=600$ MeV 
in the three dimensional momentum space 
with the coupling constant $G\Lambda^2=2.45$ \cite{MB}. 
With these parameters, the vacuum values of the pion decay constant $f_\pi$, 
the constituent quark mass $m$ and the quark condensate are obtained as 
$f_{\pi}=93.9$ MeV, $m=400$ MeV and 
$\langle\bar{u}u\rangle=\langle\bar{d}d\rangle=(-244.9{\rm MeV})^3$. 
Further discussions about the stability of 
nonstrange quark matter can be found in Refs. \cite{MB,MA}.

We now turn to discussions on finite size droplets. 
Within the framework of the MRE \cite{MRE,JM,BJ}, 
we write the density of states 
for a spherical system as $k^2\rho_{\mb{MRE}}/(2\pi^2)$, where
\begin{eqnarray}
\rho_{\mb{MRE}}=\rho_{\mb{MRE}}(k,m,R)
=1+\frac{6\pi^2}{kR}f_S\left(\frac{k}{m}\right)
+\frac{12\pi^2}{(kR)^2}f_C\left(\frac{k}{m}\right),\label{eqn:dos}
\end{eqnarray}
with $R$ being the radius of the sphere. 
The functions $f_S(k/m)$ and $f_C(k/m)$ represent the surface 
and the curvature contributions 
to the fermionic density of states, respectively. 
The functional form of the surface contribution is given by
\begin{eqnarray}
f_S\left(\frac{k}{m}\right)=-\frac{1}{8\pi}
\left(1-\frac{2}{\pi}\atan\frac{k}{m}\right).\label{eqn:fs}
\end{eqnarray}
On the other hand, the curvature contribution 
for an arbitrary quark mass has not been derived in the MRE. 
In this paper, we adopt the following {\it Ansatz} by Madsen \cite{JM}:
\begin{eqnarray}
f_C\left(\frac{k}{m}\right)=\frac{1}{12\pi^2}\left[1-\frac{3k}{2m}
\left(\frac{\pi}{2}-\atan\frac{k}{m}\right)\right]\label{eqn:fc}.
\end{eqnarray}
Note that Eqs. (\ref{eqn:fs}) and (\ref{eqn:fc}) have the 
following $m \to 0$ limits:
\begin{eqnarray}
\lim_{m \to 0}f_S(k/m)=0,~
\lim_{m \to 0}f_C(k/m)=\frac{-1}{24\pi^2}.
\end{eqnarray}
Thus, the effective potential for the spherical system which is 
regarded as a function of $m$, $\mu$ and $R$ is given by 
\begin{eqnarray}
\omega&=&\frac{m^2}{4G}
-\nu\int_0^{\Lambda}\frac{k^2dk}{2\pi^2}\rho_{\mb{MRE}}E_k\nonumber\\
&&-\nu T\int_0^{\Lambda}\frac{k^2dk}{2\pi^2}
\rho_{\mb{MRE}}\ln\left[1+e^{-\beta(E_k+\mu)}\right]
\left[1+e^{-\beta(E_k-\mu)}\right],\nonumber\\
&\underset{T \to 0}{\longrightarrow}&\frac{m^2}{4G}
-\nu\int_{k_F}^{\Lambda}\frac{k^2dk}{2\pi^2}\rho_{\mb{MRE}}E_k
-\nu\mu\int_0^{k_F}\frac{k^2dk}{2\pi^2}\rho_{\mb{MRE}},
\end{eqnarray}
where the Fermi momentum, $k_F$, is related to the chemical potential 
by $\mu=\sqrt{k_F^2+m^2}$. 
Henceforth we restrict ourselves to zero temperature. 
For computation of a finite system, we choose a fixed 
baryon number $A$ and a radius of the sphere $R$. 
The quark mass $m$ is then determined by 
the Schwinger--Dyson equation (SDE) that is 
the extremum condition of $\omega$ 
with respect to $m$: $\partial\omega/\partial m=0$. 
The resulting SDE is given by
\begin{eqnarray}
m=2G\nu\frac{\partial}{\partial m}
\int_{k_F}^{\Lambda}\frac{k^2dk}{2\pi^2}
\rho_{\mb{MRE}}E_k+2G\nu\mu\frac{\partial}{\partial m}
\int_0^{k_F}\frac{k^2dk}{2\pi^2}\rho_{\mb{MRE}}.\label{eqn:sd1}
\end{eqnarray}
In addition to the SDE, since the baryon number is fixed to $A$, 
we have another equation, i.e.
\begin{eqnarray}
Vn_B=V\frac{\nu}{3}\int_0^{k_F}\frac{k^2dk}{2\pi^2}\rho_{\mb{MRE}}=A,
\label{eqn:sd2}
\end{eqnarray}
where $V=4\pi R^3/3$ is the volume of the droplet and 
$n_B$ denotes the baryon number density in the droplet 
that is one third of the quark number density $n_q$:
\begin{eqnarray}
n_B=\frac{n_q}{3}
=-\frac{1}{3}\left(\frac{\partial\omega}{\partial\mu}\right)_T
\underset{T\to 0}{\longrightarrow}
\frac{\nu}{3}\int_0^{k_F}\frac{k^2dk}{2\pi^2}\rho_{\mb{MRE}}.
\end{eqnarray}
Equations (\ref{eqn:sd1}) and (\ref{eqn:sd2}) have to be solved 
self-consistently. Therefore, by virtue of Eq. (\ref{eqn:sd2}), 
one can reduce these equations to the following set of coupled equations:
\begin{eqnarray}
&&m=2G\nu\int_{k_F}^{\Lambda}\frac{k^2dk}{2\pi^2}
\frac{\partial}{\partial m}
\left(\rho_{\mb{MRE}}E_k\right),
\label{eqn:rsd1}\\
&&V\frac{\nu}{3}\int_0^{k_F}\frac{k^2dk}{2\pi^2}\rho_{\mb{MRE}}=A.
\label{eqn:rsd2}
\end{eqnarray}
Notice that we consider $k_F$ in Eq. (\ref{eqn:rsd1}) to be 
a variable independent of $m$. 
However, under the constraint (\ref{eqn:rsd2}) $k_F$ depends on $m$. 
Therefore, these equations are also obtained by minimizing 
the energy density
\begin{eqnarray}
\epsilon&=&\frac{m^2}{4G}
-\nu\int_{k_F}^{\Lambda}\frac{k^2dk}{2\pi^2}\rho_{\mb{MRE}}E_k
\end{eqnarray}
with the constraint of fixed baryon number.

Once we know the density dependent (i.e. the radius dependent) quark mass $m$, 
we can compute various physical quantities. 
For instance, the pressure $p$ inside the droplet are given by
\begin{eqnarray}
p=-\frac{m^2}{4G}+\nu\int_{k_F}^{\Lambda}\frac{k^2dk}{2\pi^2}
\rho_{\mb{MRE}}E_k+\frac{3A}{V}\sqrt{k_F^2+m^2}-p_{\mb{vac}},
\label{eqn:press}
\end{eqnarray}
where $p_{\mb{vac}}$ has been introduced to 
ensure $p \to 0$ when $k_F \to 0$ (i.e. $R\to\infty$). 

We note that, unlike the bulk case, the SDE (\ref{eqn:rsd1}) does not 
have a trivial solution $m=0$. 
The reason is that $\rho_{\mb{MRE}}$, 
the density of states of the MRE, 
depends on the quark mass $m$, and hence 
the r.h.s. of the SDE (\ref{eqn:rsd1}) is 
no longer proportional to $m$, in contrast to the bulk case.

\section{Numerical Results}
In this section, we present the numerical results. 
Unless stated otherwise, the baryon number $A$ is fixed to $A=100$.

For a given baryon number, a nontrivial solution 
to the coupled equations (\ref{eqn:rsd1}) and (\ref{eqn:rsd2}) is 
obtained as a function of the radius $R$. 
Then, as shown in Fig. 1, the pressure of chirally symmetric 
and broken phases are calculated from Eq. (\ref{eqn:press}). 
As a result, the critical radius at which the chiral phase transition 
occurs is identified as $R_c^{\mt{(MRE)}}\simeq 12$ fm. 
The corresponding critical baryon number density $n_c^{\mt{(MRE)}}$ 
is rather small $n_c^{\mt{(MRE)}} \simeq 0.08n_0$ with
$n_0$ being normal nuclear matter density $n_0=0.17~{\rm fm}^{-3}$.
\footnote{In the case of bulk matter, our model parameters yield 
the critical density $n_c^{\mt{(QM)}} \simeq 2n_0$.}

The resultant quark mass as a function of the radius 
is shown in Fig. 2. Since the baryon number is fixed, 
by changing the radius $R$, the baryon number density $n_B$ changes. 
Therefore, Fig. 2 implies that the phase transition 
as $n_B$ is varied is of first-order 
when the effect of the MRE is included. 
For comparison, we also present the result without the MRE 
showing that the transition at $R_c \simeq 4$ fm is of second-order. 
The difference between these two curves obviously stems from 
the finite size effects. 
This result is roughly interpreted as follows. 
In the MRE, the finite size effects reduce the density of states and, 
therefore, the {\it actual} Fermi momentum increases 
for a given baryon number. 
The increasing tendency of the Fermi momentum is more pronounced 
when the baryon number is small (see Fig. 3). 
As a consequence, the chirally broken phase is energetically 
disfavored for small baryon numbers. 

Figure 4 shows the quark mass as a function of the radius 
for the cases of $A=100$ and $A=1000$. 
It is clear that, for relatively large baryon numbers, 
the first-order transition is weakened. 
In addition, at $A=1000$ we find $R_c^{\mt{(MRE)}}\simeq 14$ fm 
and $n_c^{\mt{(MRE)}} \simeq 0.5n_0$. 
These results, off course, indicate that 
the finite size effects become less 
important for large baryon numbers. 

Let us consider the radius of the stable droplet from the 
usual pressure balance relation between the inside of the 
droplet and the vacuum (the outside of the droplet). 
The pressure inside the droplet with and without the MRE 
are shown in Fig. 5 for $A=100$, 
where the vacuum pressure is taken to be zero. 
When $R$ is small, the pressures are positive and 
chiral symmetry is restored inside the droplet. 
As $R$ is increased the pressures decrease. 
Now without the finite size effects (MRE) the pressure keeps 
decreasing in the negative pressure region toward the cusp point, 
where the chiral phase transition takes place. 
Beyond the cusp point, the pressure curve follows 
the one in the chirally broken phase and turns into 
the positive region (but with small values), 
asymptotically approaching the zero value. 
There are two points where the pressure curve passes the zero, 
where the system is in equilibrium. 
The left point ($R \simeq 3.7$ fm) corresponds 
to the absolute minimum of the energy per baryon, 
while the right point ($R \simeq 7.5$ fm) corresponds to the local maximum. 
Consequently, the stable droplet of radius $R_s\simeq 3.7$ fm 
lies in the chirally symmetric phase. 

On the other hand, with the MRE the pressure crosses zero 
only at $R\simeq 3.5$ fm. 
We have calculated the pressure up to $R=10^8$ fm and confirmed 
that the pressure of the chirally broken phase is always negative. 
Therefore, we find $R_s^{\mt{(MRE)}}\simeq 3.5$ fm, which is
somewhat small as compared to that without the MRE, 
and the stable droplet lies in the chirally symmetric phase. 
The radius of a stable droplet is presented in Fig. 6 
as a function of $A$. 
As is evident from the figure, 
as long as $A$ is sufficiently large, 
$R_s^{\mt{(MRE)}}$ varies as $A^{1/3}$. 
This result implies that, as mentioned before, the finite size effects 
are less important at large baryon numbers. 
Incidentally, we have observed 
the saturation property ($R\sim A^{1/3}$) when $A$ is large. 
This is a consequence of the balance between 
the kinetic energy of quarks and 
the (bag like) volume type potential energy.

\section{Conclusions}
We have studied the behavior of chiral symmetry 
in finite systems within the ${\rm SU(2)}_L \times {\rm SU(2)}_R$ 
symmetric Nambu--Jona-Lasinio model. 
The multiple reflection expansion was used to 
take account of finite size effects. 
We considered spherical quark droplets 
with baryon numbers $A=100-1000$. 
The Schwinger--Dyson equation for the quark mass 
was solved and, then, it turned out that the finite size effects 
enhance chiral symmetry restoration. 
The critical radii with and without the MRE differed significantly 
from each other. 
If the finite size effects  are important also for strange quarks, 
it may have significant effects on 
the (meta)stability of strangelets. 
On one hand, the enhancement of restoring chiral symmetry 
lowers the constituent mass of strange quarks. 
This affects the argument made by Buballa and Oertel \cite{BO} 
who pointed out that the large strange quark mass at the 
relevant densities is the main reason 
that makes strange quark matter unstable. 
On the other hand, the finite size effects 
increase the energy of strangelets \cite{SQM,JM,BJ,GJ}. 
These two competing aspects should be carefully treated in realistic 
calculations

In order to find the radius of the stable droplet, 
we calculated the pressure inside the droplet. 
Without the finite size effects (MRE) the pressure shows the characteristic 
behavior as shown in Fig. 5. 
Such behavior has been also obtained \cite{MA,SKP} 
with interactions/parameters which differ 
from those in this paper. 
In contrast, with the MRE its behavior changes considerably. 
We found that the pressure of the chirally broken phase 
is always negative and, therefore, 
the chirally broken phase is unstable against collapse. 
However, the radius of the stable droplet 
does not change much as compared to that without the MRE. 
After all, it was found that chiral symmetry is restored 
in the stable droplet. 
This result reminiscent of the MIT bag model 
supports the former results without the MRE \cite{MB,MA}. 

We have already referred to the validity of 
the MRE in the first section. 
It should be also noted that the MRE contains 
several problems concerning its reliability. 
In the first place, $\rho_{\mb{MRE}}$ should have 
terms proportional to $1/R^3$, $1/R^4$, and so on, 
though their functional forms are not known. 
These terms would be dominant at small radii. 
Furthermore, it is well known that the MRE causes unphysical 
negative density of states at small radii. 
Neergaard and Madsen \cite{NM} have investigated the validity of the MRE and 
proposed a solution to these problems. 
However, for systems of $A \agt 50$, these problems have 
rather minor effects and the results presented here should 
hold when an improved density of states is employed. 

In the present analysis we have adopted the simplest NJL model
with scalar and pseudoscalar channels.  
At the quantitative level the inclusion of 
vector and axial-vector channels could be important which brings repulsive 
correlations~\cite{MB,asakawa,klimt,inm}.  
It might modify the present result such that, for instance, the 
transition point would be shifted to higher density side.  
However, we expect that our main 
conclusion drawn here on the relative effect of finite cavity 
as compared to the bulk case should be rather stable.  
This is, however, an interesting question to be investigated in the 
future work.  

Finally, we comment on the outlook for future studies. 
Physics of strangelets and 
the quark droplets at finite temperature should 
have important implications for 
the cosmological quark-hadron phase transition 
and heavy ion collision experiments. 
It is worthwhile doing more detailed analysis
including charge neutrality, chemical equilibrium by weak interactions 
and the color singlet constraint. 
Furthermore, it is also interesting 
to include other dynamical effects 
as well as the chiral phase transition, e.g., color superconductivity 
and color flavor locking \cite{CSC,CFLS,CSCFS}. 
They may alter the present phase structure at small radii, 
because, at sufficiently low temperatures and high densities, 
the chirally broken phase would undergo 
a phase transition into the color superconducting phase. 

\begin{acknowledgments}
We are grateful to S. Raha for suggesting this subject 
and for stimulating discussions. We also thank to N. Sandulescu 
and S. Yasui for valuable discussions.
\end{acknowledgments}

\begin{figure}
\includegraphics{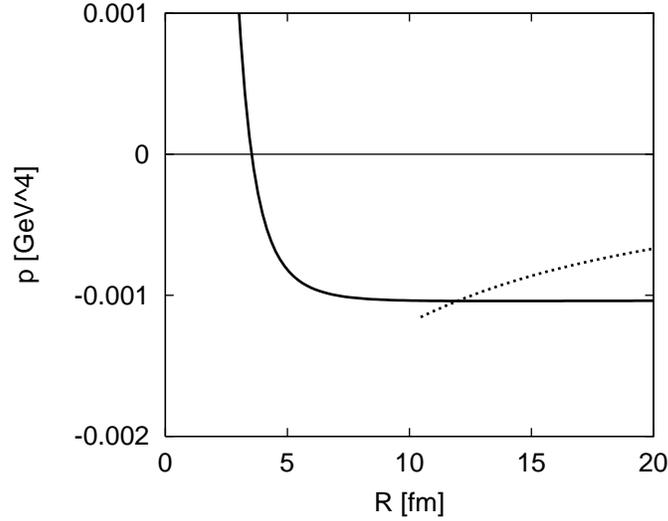}
\caption{The pressure inside a droplet as a function of $R$ 
for $A=100$. The solid line shows the pressure of chirally symmetric phase 
and the dotted line shows that of chirally broken phase. 
Two lines cross at $R \simeq 12$ fm. 
Notice that the vacuum (the outside of the droplet) 
pressure is taken to be zero.}
\end{figure}

\begin{figure}
\includegraphics{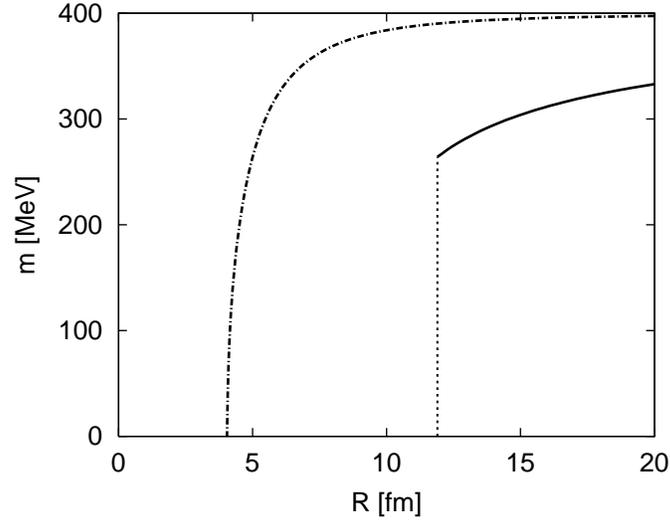}
\caption{Radius dependence of the quark mass 
for $A=100$ (solid line). For comparison, the result without the MRE 
is presented by dot-dashed line.}
\end{figure}

\begin{figure}
\includegraphics{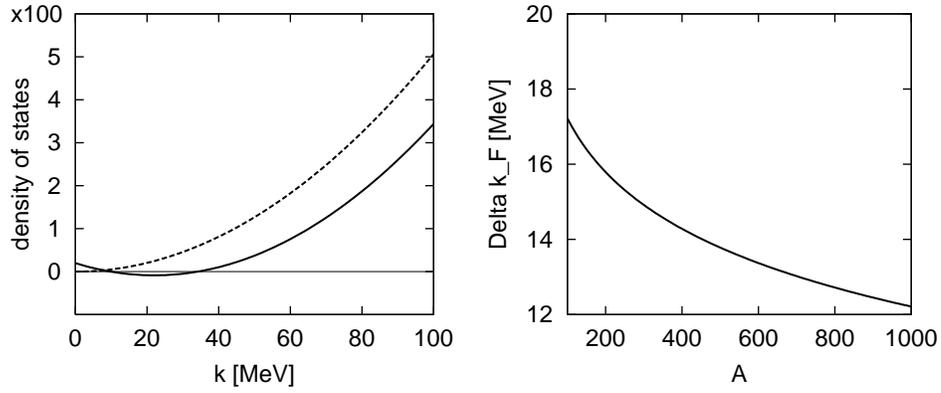}
\caption{The density of states 
$k^2\rho_{\mb{MRE}}/(2\pi^2)$ (solid line) and that 
without the MRE (dotted line) 
in unit of $\mbox{MeV}^2$ as a function of $k$ (left panel). 
The increase of the Fermi momentum due to 
the finite size effects $\Delta k_F$ as a function of $A$ (right panel). 
The quark mass $m$ and the radius $R$ are 
taken to be their typical values; $m=200$ MeV and $R=10$ fm.}
\end{figure}

\begin{figure}
\includegraphics{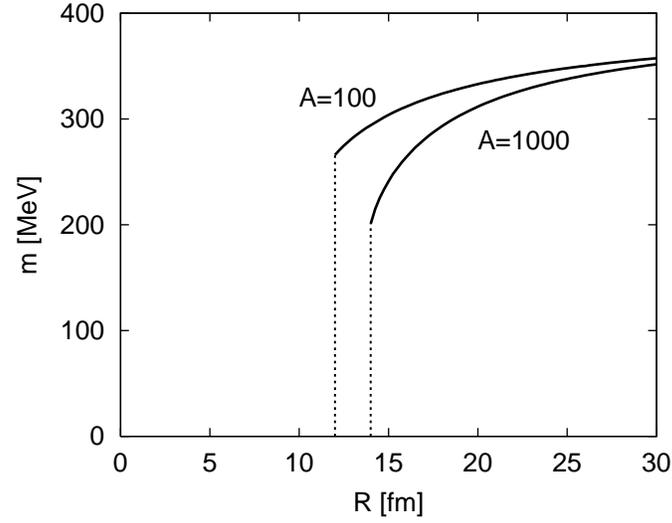}
\caption{The radius dependence of the quark mass for the cases 
of $A=100$ and $A=1000$. 
For relatively large baryon number, the first-order transition is weakened.}
\end{figure}

\begin{figure}
\includegraphics{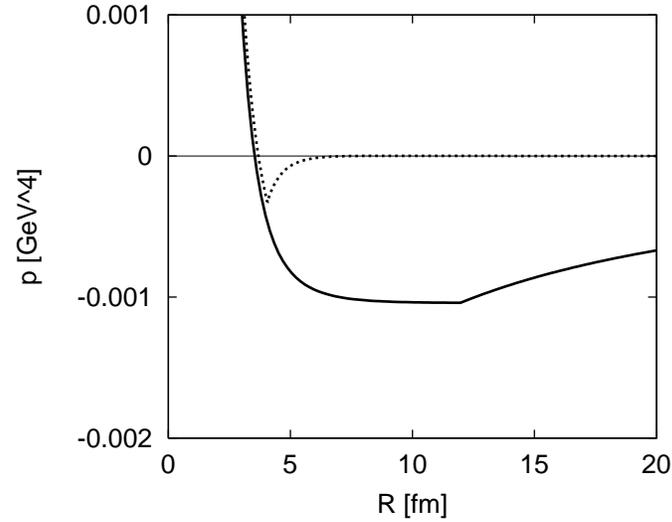}
\caption{The pressure inside a droplet ($A=100$) as a function of $R$. 
The curves correspond to the pressure with (solid line) 
and without (dotted line) the MRE. Without the MRE 
the pressure becomes positive for $R \agt 7.5$ fm, 
though it is difficult to see in this figure.}
\end{figure}

\begin{figure}
\includegraphics{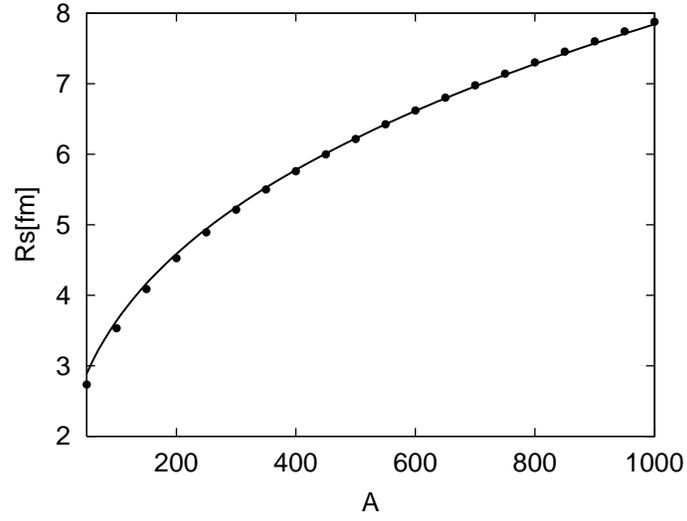}
\caption{The radius of the stable droplet $R_s^{\mt{(MRE)}}$ 
as a function of baryon number (blobs). 
The solid line shows $\chi^2$ fitting to 
$R_s^{\mt{(MRE)}}\propto A^{1/3}$.}
\end{figure}

\end{document}